\documentstyle[12pt]{article}
\catcode`@=11
\def\be{\begin{enumerate}}   \def\ee{\end{enumerate}}
\def\fo{\hbox{{1}\kern-.25em\hbox{l}}}
\def\fnote#1#2{\begingroup\def\thefootnote{#1}\footnote{#2}\addtocounter
{footnote}{-1}\endgroup}
\def\utgp{Theory Group\\ Department of Physics \\ University of Texas
\\ Austin, Texas 78712}
\catcode`@=12

\begin{document}

\hfill{UTTG-30-92}

\hfill{December 1992}

\vfill
\vspace{24pt}
\begin{center}

{\bf On the Black Hole Background of Two-Dimensional String Theory}

\vspace{24pt}

Shyamoli Chaudhuri\fnote{*}{sc@utaphy.bitnet} and
Djordje Minic\fnote{\dagger}{minic@utaphy.bitnet}

\vspace{12pt}
\utgp
\vspace{32pt}

\end{center}

\begin{abstract}
The classical black hole background of two-dimensional string theory
is examined after including the effect of the tachyon field.
Keeping all terms upto $O(T^2)$, and making no other approximations,
the only consistent classical solution to the resulting dilaton-graviton
theory is found to be flat spacetime with a nontrivial dilaton.
\end{abstract}

\pagebreak
\setcounter{page}{1}

The discovery of a nontrivial dilaton-graviton background of
critical two-dimensional string theory with the target space geometry
of a 2-d black hole \cite{eli} \cite{man} \cite{wit} has stimulated
much activity in the quantum mechanics of black holes \cite{harvey}.
Subsequently attempts have been made to include the effects of the tachyon
in the solution of the beta function equations \cite{shanta} \cite{russo}.
However these treatments have thus far been restricted to the asymptotic
regime, since the coupled equations are too difficult to handle in general,
and can only be compared to the conformal field theory in this regime
\cite{CL}.
In this paper, we will propose a different approach to including the effect
of the tachyon with a rather surprising result. We find
that upon eliminating the tachyon through its equation of motion, the two
dimensional black hole is no longer a consistent classical solution to the
resulting dilaton-graviton theory.

We follow the sigma model approach of Mandal, et al \cite{man}.
As is well known by now, two-dimensional critical string theory
can be viewed as a theory of two-dimensional quantum gravity coupled to
$c=1$ matter \cite{joe}, or in other words, as a non-critical Polyakov string
theory in a single embedding dimension $x^{1}(\xi )$.
In conformal gauge, $g_{ab}=\exp(\eta(\xi ))\hat{g}_{ab}$, the conformal field
theory on the world-sheet is a theory of two scalars, one being the embedding
dimension $x^{1}(\xi )$ and the other the Liouville field $\eta(\xi )$.
The corresponding sigma-model action is
\begin{equation}
S=\frac{1}{4\pi}\int d^{2}\xi \sqrt{\hat{g}}(\frac{1}{2}\hat{g}^{ab}
G_{\mu \nu}(x)\partial_{a}x^{\mu}\partial_{b}x^{\nu} -\hat{R}^{(2)}\Phi(x)+
T(x))
\end{equation}
where $G_{\mu \nu}$, $\Phi$ and $T$ denote the low energy excitations of
a critical two-dimensional bosonic string: the graviton, the dilaton and
the tachyon respectively. Also $x^{\mu}=(x^{1}(\xi ), \eta(\xi ))$.
The only propagating mode is the tachyon, which is massless in two dimensions.
The graviton and dilaton serve as auxiliary fields that parametrize the
background.

The target space equations of motion for these excitations are determined by
the
conditions of conformal invariance, i.e., the relevant beta-functions are set
to
zero \cite{call}. As is well-known, these equations can be derived
from the target space action
\begin{equation}
I=\int d^{2}x \exp(-2\Phi) \sqrt{G}(R-4(\nabla \Phi)^{2}+ (\nabla T)^{2}
      + V(T) + c)
\end{equation}
where $c=-8$ in two dimensions. We will assume
that the tachyon potential is quadratic, $V(T)=-2T^2$.
The corresponding equations of motion are given by
\begin{equation}
R_{\mu \nu}-2\nabla_{\mu}\nabla_{\nu}\Phi +\nabla_{\mu}T\nabla_{\nu}T=0
\end{equation}
\begin{equation}
R+4(\nabla \Phi)^{2}-4(\nabla^{2} \Phi) +(\nabla T)^{2}+V(T) +c=0
\end{equation}
\begin{equation}
-2\nabla^{2}T +4\nabla \Phi \nabla T +\dot{V}(T)=0.
\end{equation}

The black hole solution of \cite{man} is obtained by setting
the tachyon field to zero. Then equations (3) and (4) read
\begin{equation}
R_{\mu \nu} -2\nabla_{\mu}\nabla_{\nu}\Phi=0
\end{equation}
\begin{equation}
R + 4(\nabla \Phi)^{2} - 4 \nabla^{2} \Phi +c=0.
\end{equation}

We will work in the conformal gauge. Then,
\begin{equation}
ds^{2} = \exp(\sigma)dudv
\end{equation}
and the components of the spacetime metric are
$G_{uv}=\frac{1}{2}\exp(\sigma)$ and $G_{uu}=G_{vv}=0$. The only
nonvanishing components of the connection, $\Gamma^{a}_{bc}$, are
$\Gamma^{u}_{uu}=\partial_{u} \sigma, \Gamma^{v}_{vv}=\partial_{v} \sigma$.
The Ricci tensor is given by $R_{uv}=\partial_{u}\partial_{v} \sigma$
and the scalar curvature equals
$R=4\exp(-\sigma)\partial_{u}\partial_{v} \sigma$.

Mandal, et al found a non-trivial solution to eqns. (6) and (7),
\begin{equation}
\exp(-2\Phi) = 2uv +a,
\end{equation}
and
\begin{equation}
\exp(\sigma) = \frac{1}{2uv+a}
\end{equation}
which implies
\begin{equation}
ds^{2}=\frac{dudv}{2uv+a}.
\end{equation}

For $a\neq 0$ this line element represents a black hole in Kruskal
coordinates, with horizon at $uv=0$ and a curvature singularity at
$uv=-\frac{a}{2}$. The string coupling constant $g_{st}^2=\exp(2 \Phi)$
grows infinitely large at the location of the curvature singularity.

We will now take into account the dynamics of the tachyon
mode in this background. (A linearized analysis of equations (3) to (5) for
non-vanishing static tachyons, valid
in the asymptotic regime, is carried out in \cite{shanta},
and the extension to non-static tachyons is done in
\cite{russo}.) We begin by manipulating the action.
Let us scale the tachyon as follows, $t=\exp(-\Phi) T $.
Then the tachyonic action, namely
\begin{equation}
I_{T} = \int d^{2}x \exp(-2\Phi)\sqrt{G}((\nabla T)^{2} -2T^{2}),
\end{equation}
becomes
\begin{equation}
I_{t}= \int d^{2}x \sqrt{G}(G^{\mu \nu}\nabla_{\mu}t\nabla_{\nu}t -
          \frac{1}{4}Rt^{2})
\end{equation}
where we have used the classical equation of motion for the dilaton field
(4), and have dropped terms of $O(T^3)$ since we are interested in
small fluctuations of the tachyon about its quadratic maximum.
Let us eliminate $t$ by using its classical equation of motion
\begin{equation}
(\Box +\frac{1}{4}R)t = 0.
\end{equation}
The appropriate boundary condition is $t\rightarrow 1$ at infinity.
Then we can write the solution of the equation of motion:
\begin{equation}
t_{cl}= 1-\frac{1}{4}\hat{F}R
\end{equation}
where
\begin{equation}
\hat{F}^{-1} \equiv \Box +\frac{1}{4}R.
\end{equation}
Note that by using the equation of motion one can express the
action $I_{t}$ as a boundary term
\begin{equation}
I_{t} = \int d^{2}x \sqrt{G} \Box t_{cl}
\end{equation}
or using equation (15)
\begin{equation}
I_{t} = -\frac{1}{4} \int d^{2}x \sqrt{G}R +
          \frac{1}{16} \int d^{2}x \sqrt{G} R \hat{F} R.
\end{equation}
This procedure is familiar from the work of Fradkin and Vilkovisky on four-
dimensional quantum gravity \cite{grisha}.

In conclusion we obtain the following "effective" action for the coupled
graviton-dilaton system:
\begin{eqnarray}
I_{eff} &=& \int d^{2}x \exp(-2\Phi) \sqrt{G} (R -4(\nabla \Phi)^{2})
            - \frac{1}{4} \int d^{2}x \sqrt{G} R  \nonumber \\
        & & +\frac{1}{16} \int d^{2}x \sqrt{G} R\frac{1}{\Box +\frac{1}{4}R}R.
\end{eqnarray}
The non-local piece can be evaluated formally as an expansion in powers of
$ \Box^{-1} R $. In the following, we will approximate
$\frac{1}{\Box +\frac{1}{4}R}$ by the leading term in this expansion.
The consistency of this approximation will be justified by the result.

By varying $I_{eff}$ we obtain the relevant equations of motion for the
graviton dilaton system.
First we recall that the Einstein equations are identically satisfied in two
dimensions. Also, the variation of $W=\int d^{2}x R\frac{1}{\Box}R$ is well
known to be given by (see for example \cite{grisha1})
\begin{eqnarray}
\delta W &=& -\sqrt{G}\delta G_{\mu \nu}(2\nabla^{\mu}\nabla^{\nu}(\hat{O}R) +
             \nabla^{\mu}(\hat{O}R)\nabla^{\nu}(\hat{O}R) \nonumber \\
         & & -G^{\mu \nu}[2R +
             \frac{1}{2}\nabla_{\xi}(\hat{O}R)\nabla^{\xi}(\hat{O}R)])
\end{eqnarray}
where $\hat{O}\equiv \frac{1}{\Box}$. (In the conformal gauge
$\Box=4\exp(-\sigma) \partial_{u}\partial_{v}$, so $\hat{O}R=\sigma$.)

Hence we are left with the following set of equations:
\begin{eqnarray}
\partial_{u}\partial_{v}\sigma &=& 2\partial_{u}\partial_{v}\Phi +\frac{b}{2}
                                   (-3\partial_{u}\partial_{v}\sigma
                                    +\partial_{u} \sigma\partial_{v}\sigma)
                                    \exp(2\Phi).
\end{eqnarray}
\begin{equation}
2(\partial_{u}^{2}\Phi-\partial_{u}\sigma\partial_{u}\Phi) = -\frac{b}{2}
(2\partial_{u}^{2}\sigma-(\partial_{u}\sigma)^{2})\exp(2\Phi)
\end{equation}
\begin{equation}
2(\partial_{v}^{2}\Phi-\partial_{v}\sigma\partial_{v}\Phi) = -\frac{b}{2}
(2\partial_{v}^{2}\sigma-(\partial_{v}\sigma)^{2})\exp(2\Phi)
\end{equation}
\begin{equation}
\partial_{u}\partial_{v}\sigma +4\partial_{u}\Phi\partial_{v}\Phi-
4\partial_{u}\partial_{v}\Phi = 2\exp(\sigma) .
\end{equation}
Here $b=\frac{1}{8}$. (Of course if we set $b=0$ we recover the equations of
the
graviton-dilaton system with $T=0$.)

The goal is now to check whether the black hole metric satisfies the
improved equations of
motion derived from the effective action. Note that since we have left the
dilaton field unspecified thus far, an {\it arbitrary} two dimensional black
hole metric can, through a coordinate transformation, be brought to the
form (10). With no loss of generality then, we can substitute equation (10)
in this set of equations. We get
\begin{equation}
\partial_{u}\partial_{v}\Phi=-\frac{a}{(2uv+a)^{2}}-\frac{b}{4}
\frac{6a+4uv}{(2uv+a)^{2}}\exp(2\Phi)
\end{equation}
\begin{equation}
(\partial_{u}^{2}\Phi+\frac{2v}{2uv+a}\partial_{u}\Phi)=-\frac{b}{4}
\frac{4v^{2}}{(2uv+a)^{2}}\exp(2\Phi)
\end{equation}
\begin{equation}
(\partial_{v}^{2}\Phi+\frac{2u}{2uv+a}\partial_{v}\Phi)=-\frac{b}{4}
\frac{4u^{2}}{(2uv+a)^{2}}\exp(2\Phi)
\end{equation}
\begin{equation}
\partial_{u}\Phi \partial_{v}\Phi -\partial_{u}\partial_{v}\Phi=
\frac{uv+a}{(2uv+a)^{2}}.
\end{equation}
These equations turn out to be consistent only if $a=0$.
This is seen by eliminating $\exp(2\Phi)$ from equations (26) and (27)
which implies
\begin{equation}
u^{2}(\partial_{u}^{2}\Phi +\frac{2v}{2uv+a}\partial_{u}\Phi)=v^{2}
(\partial_{v}^{2}\Phi+\frac{2u}{2uv+a}\partial_{v}\Phi)
\end{equation}
and then substituting in equation (25) which implies
\begin{equation}
\partial_{u}\partial_{v}\Phi=\frac{v}{u}\partial_{v}^{2}\Phi +
\frac{1}{u}\partial_{v}\Phi
\end{equation}
and it therefore follows that $a=0$.

Now if $a=0$ the equations (25) to (28) read
\begin{equation}
\partial_{u}\partial_{v}\Phi = -\frac{b}{4uv}\exp(2\Phi)
\end{equation}
\begin{equation}
\partial_{u}^{2}\Phi +\frac{1}{u}\partial_{u}\Phi =-\frac{b}{4u^{2}}\exp(2\Phi)
\end{equation}
\begin{equation}
\partial_{v}^{2}\Phi +\frac{1}{v}\partial_{v}\Phi =-\frac{b}{4v^{2}}\exp(2\Phi)
\end{equation}
\begin{equation}
\partial_{u}\Phi\partial_{v}\Phi -\partial_{u}\partial_{v}\Phi = \frac{1}{4uv}.
\end{equation}

It is easily seen that the solution of the last equation is given by
\begin{equation}
\exp(-\Phi) = A\exp(C_{1}\ln \sqrt{u}+\frac{1}{C_{1}}\ln \sqrt{v}) +
B\exp(C_{2}\ln \sqrt{u} +\frac{1}{C_{2}}\ln \sqrt{v}).
\end{equation}
By subsituting this expression in (31), (32) and (33) we infer that
 $AB=b=\frac{1}{8}$ and $C_{1}=-C_{2}=1$.

Therefore the improved classical equations of motion yield flat spacetime
with the dilaton field given by
\begin{equation}
\Phi = -\ln (A\sqrt{uv}+\frac{b}{A}\frac{1}{\sqrt{uv}})
\end{equation}
while the tachyon asymptotically ($t \to 1 $) approaches
\begin{equation}
T =\exp(\Phi)= \frac{1}{A\sqrt{uv}+\frac{b}{A}\frac{1}{\sqrt{uv}}}
\end{equation}
and we recall that $b=\frac{1}{8}$.

By formally letting $b\rightarrow 0$, and comparing to the $T=0$ solution
we determine $A=\sqrt{2}$. Our solution again has the property
that the string coupling constant
$g_{st}^{2}=\exp(2\Phi)$ grows infinitely large at
$uv=-\frac{b}{A^{2}}=-\frac{1}{16}$.

What is the implication of our result? It appears that the target space black
hole background of two-dimensional string theory could be unstable
when the dynamics of the only propagating field in the theory is taken
into account, or in other words this background does not represent a true
vacuum for the graviton-dilaton system once the nonlinear
dilaton-tachyon and graviton-tachyon couplings are considered.
(This result was known to Sasha Polyakov \cite{sasha}.) Note that these
couplings are not probed in the linearized tachyon equation solved
in \cite{shanta} \cite{russo}. Our calculation neglects higher order
tachyon self-interactions as well as couplings to the massive modes of
the string. It is possible that these are essential to the stability of the
black hole solution, as suggested by the conformal field theory analysis
\cite{CL}.

What we find particularly intriguing is the appearance of a non-local term
proportional to $\int d^{2}x \sqrt{G} R\frac{1}{\Box +\frac{1}{4}R}R$ in
the spacetime effective action for the graviton-dilaton system. This term is of
a similar form to the one proposed years ago by Fradkin and Vilkovisky which
when added to the usual Einstein-Hilbert term would represent the
appropriate quantum action for four-dimensional quantum gravity (with manifest
off-shell conformal invariance) and
would yield identical on-shell properties as the Einstein theory
 (for more details consult \cite{grisha})! Of course, from
the point of view of string theory, this term is a logical possibility
in the low-energy effective action that should account for the infrared
properties of the gravitational field. In view of recent activity on the
problem of black hole evaporation \cite{lenny} it is perhaps
worthwhile reconsidering
the role of such non-local terms in the gravitational effective
action.

{\bf Acknowledgements}

We thank W.Fischler, J.Lykken, and J.Polchinski for discussions.
Our interest in this problem was considerably influenced by lectures
delivered by L.Susskind on black hole evaporation at the University of Texas.
D.M. would like to acknowledge many illuminating conversations with
Sasha Polyakov about the black hole problem in general. This work is supported
in part by the Robert A.Welch Foundation, NSF grant PHY 9009850, and the
Texas Advanced Research Program.

\pagebreak

\end{document}